\documentclass[reprint]{revtex4-2}

\usepackage{xcolor}
\usepackage[british]{babel} 
\usepackage{amssymb,amsmath,bm} 
\usepackage[ansinew]{inputenc} 
\usepackage{epsfig} 
\usepackage{subfigure}
\usepackage{tikz}
\usepackage{pgf}
\usepackage{caption}
\usetikzlibrary{shapes,arrows,arrows.meta,positioning,calc}
\tikzset{
    line/.style = {
        draw,
        -{Latex[length=20pt, width=6pt]} 
    },
    cloud/.style = {
        draw, ellipse, node distance=2.5cm,
        minimum height=2em
    }
}

\begin{document}

\title{Reproducing EPR correlations without superluminal signalling: backward conditional probabilities and Statistical Independence}
\author{Simon Friederich}
\email{s.m.friederich@rug.nl}
\affiliation{University of Groningen, University College Groningen, Hoendiepskade 23/24, 9718BG Groningen, the Netherlands}

\begin{abstract}
Bell's theorem states that no model that respects \emph{Local Causality} and \emph{Statistical Independence} can account for the correlations predicted by quantum mechanics via entangled states. This paper proposes a new approach, using backward-in-time conditional probabilities, which relaxes conventional assumptions of temporal ordering while preserving \emph{Statistical Independence} as a ``fine-tuning'' condition. It is shown how such models can account for EPR/Bell correlations and, analogously, the GHZ predictions while nevertheless forbidding superluminal signalling.
\end{abstract}
\maketitle

\section{Introduction}

Bell's theorem \cite{belltheorem,chsh,bell_sep,bell_scholarpedia} demonstrates that no theory satisfying two seemingly natural conditions---\emph{Local Causality} and \emph{Statistical Independence}---can reproduce the correlations predicted by quantum theory using entangled states. We refer to these here as ``EPR correlations'' \cite{epr}. This result poses a dilemma for dynamical models aiming to solve the quantum measurement problem: how can such models accommodate EPR correlations without violating Bell's constraints?

In this paper, I propose a novel approach that introduces \emph{backward-in-time conditional probabilities} while maintaining \emph{Statistical Independence}, understood as the absence of correlations between measurement settings and past configurations. Although these conditional dependencies formally point ``backwards,'' because Statistical Independence holds they do not permit one to affect the past by manipulating present or future settings. In that sense, in contrast with other approaches that explicitly embrace ``retrocausality'' (see \cite{friederichevans} for an overview) no real retrocausality occurs in the sense of controllable backward-in-time influences. As a further consequence of Statistical Independence being preserved the model both reproduces the EPR correlations and avoids superluminal signalling, a balancing act shown to be tricky in \cite{woodspekkens}.

The discussion is organized as follows. Section 2 briefly revisits Bell's theorem and its core assumptions, using \emph{directed acyclic graphs} (DAGs) informally as a tool to illustrate probabilistic dependences. Section 3 explains why one might consider backward-in-time conditional probabilities and addresses natural objections. Section 4 develops a concrete model for reproducing the EPR correlations, which is extended in Section 5 to the Greenberger-Horne-Zeilinger (GHZ) scenario. Finally, Section 6 concludes with limitations and directions for future research.

\section{Brief review of Bell's theorem}

Bell's theorem applies to the meausrement statistics of two subsystems. We consider two measurement settings $\alpha_1$, $\alpha_2$ and corresponding outcomes $a_1, a_2$. An additional variable $\lambda$ accounts for correlations between the outcomes. Bell's theorem relies on two assumptions:

\emph{Local Causality} (LC):
\begin{eqnarray}
(LC)\quad P(a_1,a_2 \mid \lambda,\alpha_1,\alpha_2)
=
P(a_1 \mid \lambda,\alpha_1)\,P(a_2 \mid \lambda,\alpha_2),\nonumber\\
\end{eqnarray}
and \emph{Statistical Independence} (SI), also called \emph{measurement independence} or \emph{setting independence}:
\begin{eqnarray}
{\rm (SI)}\quad 
P(\lambda \mid \alpha_1,\alpha_2) = P(\lambda).
\end{eqnarray}
Here, $\lambda$ is meant to describe the relevant past conditions (or ``hidden variables'') for both measurement events. LC expresses the idea that no influence can propagate faster than light or backward in time, so each outcome depends only on the local setting and $\lambda$. SI encodes the idea that there are no ``conspiratorial'' correlations \cite{bell_scholarpedia,bell_sep} between $\lambda$ and the freely chosen measurement settings $\alpha_1,\alpha_2$.

Under LC and SI, the joint distribution factorizes:
\begin{eqnarray}
P(a_1,a_2,\lambda \mid \alpha_1,\alpha_2)
=
P(\lambda)\,P(a_1 \mid \lambda,\alpha_1)\,P(a_2 \mid \lambda,\alpha_2).\nonumber\\
\label{Bell_factorize}
\end{eqnarray}

Such a factorization can be visualized with a directed acyclic graph (DAG), as shown in Fig.~\ref{fig:bellDAG}. Each node in the DAG represents a variable, and an arrow from one node to another indicates a direct functional or probabilistic dependence. For instance, the arrow from $\lambda$ to $a_1$ reflects that $a_1$ depends on $\lambda$, while the arrow from $\alpha_1$ to $a_1$ shows that $a_1$ also depends on the local measurement setting. Because there is no arrow from $\alpha_1$ or $\alpha_2$ to $\lambda$, we recover $P(\lambda \mid \alpha_1,\alpha_2) = P(\lambda)$, and because $a_1$ and $a_2$ depend only on local inputs, we get the product of conditional probabilities in \eqref{Bell_factorize}.

\pagebreak
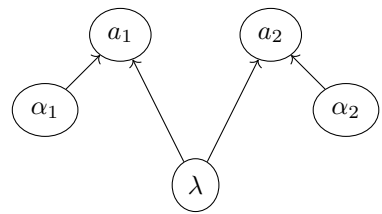
\begin{center}
\begin{tikzpicture}[node distance = 0.5cm, auto]
  \node[cloud] at (1, 0)  (x1) {$a_1$};
  \node[cloud] at (3, 0)  (x2) {$a_2$}; 
  \node [cloud] at (0,-1) (alpha1) {$\alpha_1$};
  \node [cloud] at (4,-1) (alpha2) {$\alpha_2$};
  \node [cloud] at (2,-2) (lambda) {$\lambda$};

  \draw [<-] (x1) -- (lambda);
  \draw [<-] (x2) -- (lambda);
  \draw [->] (alpha1) -- (x1);
  \draw [->] (alpha2) -- (x2);
\end{tikzpicture}
\captionsetup{font=footnotesize}
\captionof{figure}{\label{fig:bellDAG}%
DAG depicting a model in which LC and SI hold.}
\end{center}

Bell's theorem states that any model with this factorization must obey the Bell-CHSH inequality \cite{belltheorem,chsh,bell_sep,bell_scholarpedia}. Suppose $a_1, a_2 \in \{ -1, +1\}$, and we define the expectation values $\langle a_1 a_2\rangle_{\alpha_1,\alpha_2}$. Then, for two alternative measurement choices on each side ($\alpha_1,\alpha'_1$ and $\alpha_2,\alpha'_2$), the Bell-CHSH inequality is:
\begin{eqnarray}
\Big|\langle a_1a_2\rangle_{\alpha_1,\alpha_2}
- \langle a_1a_2\rangle_{\alpha_1,\alpha'_2}\Big|
&+& \Big|\langle a_1a_2\rangle_{\alpha'_1,\alpha_2}
+ \langle a_1a_2\rangle_{\alpha'_1,\alpha'_2}\Big|\nonumber\\
&\le& 2.
\label{inequality}
\end{eqnarray}
Quantum mechanics predicts violations of \eqref{inequality} using entangled states. For instance, the four standard Bell states (or EPR pairs) can be written in the $\pm 1$ basis as:
\begin{eqnarray}
|\psi_{1;2}\rangle
= \tfrac{1}{\sqrt{2}}\bigl(|1,\,1\rangle \pm |-1,\,-1\rangle\bigr),\\
|\psi_{3;4}\rangle
= \tfrac{1}{\sqrt{2}}\bigl(|1,\,-1\rangle \pm |-1,\,1\rangle\bigr).
\end{eqnarray}
In measurements of spin or polarization, restricting to coplanar measurement directions $\alpha_1,\alpha_2$, these states yield probabilities
\begin{eqnarray}
P_{1;2}(a_1,a_2 \mid \alpha_1,\alpha_2)
= \tfrac{1}{4}\bigl(1 \pm a_1 a_2\,\cos(\alpha_1-\alpha_2)\bigr),
\label{EPRB1}\\
P_{3;4}(a_1,a_2 \mid \alpha_1,\alpha_2)
= \tfrac{1}{4}\bigl(1 \mp a_1 a_2\,\cos(\alpha_1+\alpha_2)\bigr),
\label{EPRB2}
\end{eqnarray}
whose correlations can violate \eqref{inequality} for suitable angles. Therefore, no model that satisfies both (LC) and (SI) can reproduce all the predictions of quantum theory for such states.

\section{Motivating backward arrows}

In Fig.\ 1, all arrows run bottom-to-top, suggesting a forward-in-time causal direction from past to future. This ``forward'' orientation aligns with everyday thermodynamic experience.  However, neither the dynamical equations of classical mechanics nor those of quantum mechanics (the Schr\"odinger and von Neumann equations) privilege a particular time direction in a similar way (see \cite{roberts} for a recent detailed exploration of this issue). The directedness of relations in DAG-based models may hence be seen as in tension with quantum theory's qualitatively symmetric treatment of time.

One viable response to this observation is  to explore which DAG-based models, unlike those in Fig.\ 1, could recover the  EPR correlations by not restricting themselves to forward-in-time-directed arrows. So, let us consider DAGs with ``backward-in-time'' arrows into $\lambda$. Because the two wings are operationally equivalent, we focus on symmetric models. Moreover, since $\alpha_1$ and $\alpha_2$ are freely chosen measurement settings, we exclude any models where arrows \emph{into} $\alpha_1$ or $\alpha_2$ appear. These considerations leave two classes of DAGs, shown in Figs.\ 2 and~3, which differ in the direction of arrows between $a_1,a_2$ and $\lambda$. Models that contain only two arrows on each wing can be seen as special cases.



\begin{center}
\begin{tikzpicture}[node distance = 0.5cm, auto]
 \node[cloud] at (1, 0)  (x1) {$a_1$};
    \node[cloud] at (3, 0)  (x2) {$a_2$}; 
\node [cloud] at (0,-1) (alpha1) {$\alpha_1$};
\node [cloud] at (4,-1) (alpha2) {$\alpha_2$};
\node [cloud] at (2,-2) (lambda) {$\lambda$};

\draw [<-] (x1) -- (lambda);  
\draw [<-] (x2) -- (lambda);  
\draw [->] (alpha1) -- (lambda);  
\draw [->] (alpha2) -- (lambda);  
\draw [->] (alpha1) -- (x1);  
\draw [->] (alpha2) -- (x2);  
\end{tikzpicture}
\captionsetup{font=footnotesize}
\captionof{figure}{DAG with backward arrows, first type}
\end{center}

\begin{center}
\begin{tikzpicture}[node distance = 0.5cm, auto]
 \node[cloud] at (1, 0)  (x1) {$a_1$};
    \node[cloud] at (3, 0)  (x2) {$a_2$}; 
\node [cloud] at (0,-1) (alpha1) {$\alpha_1$};
\node [cloud] at (4,-1) (alpha2) {$\alpha_2$};
\node [cloud] at (2,-2) (lambda) {$\lambda$};

\draw [->] (x1) -- (lambda);  
\draw [->] (x2) -- (lambda);  
\draw [->] (alpha1) -- (lambda);  
\draw [->] (alpha2) -- (lambda);  
\draw [->] (alpha1) -- (x1);  
\draw [->] (alpha2) -- (x2);  
\end{tikzpicture}
\captionsetup{font=footnotesize}
\captionof{figure}{DAG with backward arrows, second type}
\end{center}

Various models corresponding to Fig.\ 2 have been considered in the literature \cite{brans,hall,evans}. In contrast, models corresponding to  Fig.\ 3 seem to have been largely ignored so far, perhaps because, at first glance, they do not seem to predict \emph{any} correlations between the outcomes $a_1$ and $a_2$, let alone one's that violate the Bell-CHSH inequality. This can be seen by observing that  every path connecting $a_1$ and $a_2$ passes through $\lambda$, which acts as a ``collider'' (a node with multiple incoming arrows). Normally, such a collider on a path \emph{blocks} correlation between $a_1$ and $a_2$. Hence these outcomes typically end up uncorrelated and cannot violate the Bell-CHSH inequality.

However, a suggestion by Price and Wharton \cite{pricewharton1,pricewharton2}, based on the observation that \emph{conditioning} on a collider \emph{induces} correlations that are otherwise absent $\lambda$ \cite{pearl,elwertwinship}, suggests a way around this. If, as part of the preparation procedure, $\lambda$ is set to a particular value $\lambda_0$ by the experimenters, then the Bell-correlations between $a_1$ and $a_2$ might be recoverable as artifacts of such ``collider bias.''  This re-opens the possibility of obtaining  Bell-type correlations from models with a Fig.\ 3 structure, namely, by conditioning on $\lambda_0$.  From the point of view of causal analysis, this conditioning on $\lambda_0$ amounts to ``\emph{postselection} on an early-time variable.'' Indeed, Price and Wharon also consider scenarios where the Bell-correlations are susceptible to an interpretation in terms of ordinary postselection on a late-time variable, namely scenarios with delayed choice entanglement swapping. However, in the scenarios considered in this paper, where the entanglement arises from early-time preparation of a Bell state, there is no late-time variable that would seem to be a plausible candidate for being the $\lambda_0$ with respect to which postselection is performed. Accordingly, it seems more promising to interpret the law-like probabilistic dependences indicated by the backward arrows in Figs.\ 2 and~3 as backward-in-time. 

Understandably, many have principled  concerns about backward arrows as in Figs.\ 2 and~3.  By the standards of the nowadays dominant account of causation in contemporary philosophy of science, namely, \emph{interventionism} \cite{woodward}, such models may seem to be inevitably retrocausal, and retrocausality appears inherently problematic to many. According to interventionism, a variable $X$ is a cause of some other variable $Y$ just in case intervening on $X$ is an effective means of intervening on $Y$. Since the measurement settings $\alpha_1$ and $\alpha_2$ can be manipulated (``intervened on'') and since \emph{Statistical Independence} (SI) fails in these models---namely, in general, $P(\lambda \mid \alpha_1, \alpha_2) \neq P(\lambda)$---the backward arrows from these settings into $\lambda$ indeed make models corresponding to Figs.\ 2 and~3 generally retrocausal by interventionist standards. It is worth noting, though, that the backward arrows from the \emph{outcomes} $a_1$ and $a_2$ into $\lambda$ do not per se constitute retrocausality in the interventionist sense because, contrary to the settings $\alpha_1$ and $\alpha_2$, the outcomes cannot be manipulated by experimenters. 

Postulating retrocausality that counts as such by interventionist standards can give rise to consistency challenges that arise from ``inconsistent loops'': If it is sometimes possible to (indirectly) intervene on what happened in the past, then it seems in principle possible to prevent the past causes of one's own very actions from occurring. The famous ``grandfather paradox'' epitomizes this difficult: If retrocausality were unrestrictedly possible, one could exploit it to kill one's own grandfather, thereby undermining a necessary cause of one's own existence. The scope of retrocausality needs to be carefully restricted in order to avoid inconsistencies such as these, for discussions by philosophers see \cite{lewis,price,evans,friederichevans}. It should be noted that law-like backwards-in time dependences by themselves do not give rise to these as long as they are not exploitable for manipulating the past.

Finally, one should mention a general challenge to any attempt of accounting for the Bell correlations using models that can be represented by DAGs, including ones with backward arrows, a challenge due to a result by Wood and Spekkens \cite{woodspekkens}. According to that result,  \emph{any} model reproducing EPR correlations,  including ones with backward arrows,  must violate a core assumption of causal discovery algorithms called \emph{Faithfulness}. This assumption states that any  absence of correlations between variables  in the model arises from a genuine absence of causal connections, rather than from \emph{fine-tuning} (i.e.\ delicate cancellations between arrows that would normally induce correlations). In the absence of such fine-tuning, any attempt to reproduce Bell-inequality violations inevitably introduces superluminal signaling---again contradicting quantum theory. One may interpret these findings as suggesting that a more fundamental theory than standard quantum mechanics will be ``all-at-once'',  as Wharton puts it  \cite{wharton},  i.e. that the overall cosmic history cannot be obtained by some dynamical generation procedure but only as a whole, subject to certain constraints. Such a theory plausibly could  not be formulated in terms of DAG-representable probability distributions that obey the usual rules of causal discovery.

Arguably, it is still worthwhile to investigate to what extent fine-tuned models can account for the Bell correlations, perhaps mostly for heuristic purposes, to explore the prospects for compelling single world-realist solutions to the quantum measurement problem. Yet we seem to  face a dilemma with respect to Bell's theorem. On one hand, acknowledging quantum theory's (qualitative) time symmetry motivates exploring ``backward arrow'' models. On the other hand, violating SI to enable these backward arrows raises both conceptual challenges (retrocausality and paradoxes) and the practical worry that we see no empirical sign of such backward effects. Moreover, we still confront the Wood--Spekkens result that reproducing the Bell correlations without superluminal signaling must involve \emph{fine-tuning} and thus break \emph{Faithfulness}.

\section{Statistical Independence as a fine-tuning condition}

The previous section posed a dilemma: backward-in-time arrows (as in Fig.\ 3)  offer a still underexplored route to towards recovering violations of the Bell-CHSH inequality, but they  naturally conflict with Statistical Independence (SI), which is independently plausible.  Meanwhile, the Wood--Spekkens result \cite{woodspekkens} implies that any model reproducing EPR correlations without superluminal signaling must violate \emph{Faithfulness}. This section proposes to use SI \emph{itself} as just such a fine-tuning condition. That is, we treat SI as an independently plausible assumption that---by violating Faithfulness---blocks signaling yet still recovers the quantum correlations. Preserving SI allows one to avoid retrocausality in the interventionist sense and, with it, the consistency issues mentioned in the previous section.

A natural objection  to imposing SI as a fine-tuning condition is that it  effectively removes the backward arrows from $\alpha_1,\alpha_2$ to $\lambda$ in Figs.\ 2 or~3, and thus loses the ability to replicate EPR correlations. Indeed, for Fig.\ 2, restoring SI collapses the model into Fig.\ 1, precisely the class of models that Bell's theorem rules out. In Fig.\ 3, however, imposing SI \emph{does not} simply cut those arrows. It merely requires that, upon summing over $a_1,a_2$, the distribution of $\lambda$ remain independent of $\alpha_1,\alpha_2$, even though $\lambda$ may still depend on $(a_1,a_2,\alpha_1,\alpha_2)$.

Concretely, in Fig.\ 3, we assume
\begin{eqnarray}
P(a_1,a_2, \lambda|\alpha_1,\alpha_2)=P(a_1|\alpha_1)\,P(a_2|\alpha_2)\,P(\lambda|a_1,a_2,\alpha_1,\alpha_2)\,.\nonumber\\\label{put}
\end{eqnarray}

Imposing SI means
\begin{eqnarray}
P(\lambda)
&=&P(\lambda \mid \alpha_1,\alpha_2)\nonumber\\
&=&\sum_{a'_1,a'_2}
  P(a'_1 \mid \alpha_1)\,P(a'_2 \mid \alpha_2)\,
  P(\lambda \mid a'_1,a'_2,\alpha_1,\alpha_2),\nonumber
\end{eqnarray}
with no net dependence on $\alpha_1,\alpha_2$. For the Bell states $|\psi\rangle_{1;2;3;4}$ specifically, one must choose $P(a_1=\pm1\mid \alpha_1)=P(a_2=\pm1\mid \alpha_2)=\tfrac12$.  It should be noted that this in some sense ``disables'' the arrows from the settings to the outcomes in Fig.\ 3. (However, simply omitting these arrows also does not seem attractive as it may seem to suggest, erroneously, that the outcomes $a_1$ and $a_2$ are adjustable parameters whose values can be freely fixed.) One may interpret this observation as indicating that Fig.\ 3 has mostly a heuristic role in obtaining the present model and should not be viewed as a sacrosanct enshrinement of causal structure. In any case,  the conditional probabilities $P(a_1=\pm1\mid \alpha_1)$ and $ P(a_2=\pm1\mid \alpha_2)$ would have to be adjusted if one were to extend the present approach to non-maximally entangled states.

To reproduce the quantum correlations, we further set
\begin{eqnarray}
P(\lambda_{1;2;3;4}|a_1,a_2,\alpha_1,\alpha_2)=\mathcal N P_{1;2;3;4}(a_1,a_2|\alpha_1,\alpha_2)\nonumber\\\label{recover}
\end{eqnarray}
where $P_{1;2;3;4}$ is the desired quantum probability (e.g.\ from Eqs.\ (\ref{EPRB1})--(\ref{EPRB2})) and where $\lambda_i$ labels which Bell state is prepared. (One may object here that taking $\lambda$ to label the prepared quantum state fits badly with the fact that all ``causal'' arrows into $\lambda$ come from the future in Fig.\ 3, in tension with the idea of state preparation in the past. This concern is valid and will be briefly addressed in Section 6.) Summing over $\lambda_i$ yields a normalization condition ensuring
\(\sum_i P(\lambda_i\mid a_1,a_2,\alpha_1,\alpha_2) = 1\).

Under these assignments, one recovers exactly the quantum predictions by conditioning on a particular $\lambda_i$. For instance,
\begin{eqnarray}
&&P(a_1,a_2|\alpha_1,\alpha_2,\lambda_{1})=\frac{P(a_1,a_2,\lambda_{1}|\alpha_1,\alpha_2)}{P(\lambda_{1}|\alpha_1,\alpha_2)}\nonumber\\
&&\hspace{0.5cm}=\frac{P(\lambda_{1}|a_1,a_2,\alpha_1,\alpha_2)\,P(a_1|\alpha_1)\,P(a_2|\alpha_2)}{P(\lambda_{1})}\label{success}\\
&&\hspace{0.5cm}=\frac{P(\lambda_{1}|a_1,a_2,\alpha_1,\alpha_2)}{\mathcal N}=
 P_{1}(a_1,a_2|\alpha_1,\alpha_2)\,,\nonumber
\end{eqnarray}
where we used Eq.\ (\ref{put}) in the second line, Eq.\ (\ref{recover}) in the third line, and fix $\mathcal{N}$ so that $P(\lambda_1) / [P(a_1|\alpha_1)P(a_2|\alpha_2)] = \mathcal{N}$. An analogous calculation shows the same works for $\lambda_2,\lambda_3,\lambda_4$.

Because SI is imposed, the Bell-CHSH inequality can only be violated by rejecting \emph{Local Causality} (LC). Indeed (note the inequality sign in the third line),
\begin{eqnarray}
&&\hspace{0.5cm}P(a_1|\alpha_1,\lambda_{1})\cdot P(a_2|\alpha_2,\lambda_{1})\nonumber\\
&&=\hspace{0.5cm}\frac{P(\lambda_{1}|a_1,\alpha_1)\,P(\lambda_{1}|a_2,\alpha_2)\,P(a_1|\alpha_1)\,P(a_2|\alpha_2)}{(P(\lambda_{1}))^2}\,\nonumber\\
&&\neq\hspace{0.5cm}\frac{P(\lambda_{1}|a_1,a_2,\alpha_1,\alpha_2)\,P(a_1|\alpha_1)\,P(a_2|\alpha_2)}{P(\lambda_{1})}\\
&&=\hspace{0.5cm}P(a_1,a_2|\alpha_1,\alpha_2,\lambda_{1})\,.\nonumber
\end{eqnarray}
Thus, these backward-in-time models are definitely \emph{not} local hidden-variable theories in the traditional Bell sense---rather, they break \emph{Local Causality} to evade Bell's theorem.

It is instructive to see how SI acts as a fine-tuning that \emph{prevents} superluminal signaling. Since we are considering a regime where one must condition on $\lambda$, the condition of \emph{no superluminal signalling} reads
\begin{eqnarray}
P(a_1|\alpha_1,\alpha_2,\lambda)= P(a_1|\alpha_1,\alpha'_2,\lambda)\,,\nonumber
\end{eqnarray}
for any specific value of $\lambda$.

Unless SI is implemented, this condition is generally violated in models represented by Fig.\ 3. For example, if one replaces Eq.\ (\ref{recover}) by
\begin{eqnarray}
P(\lambda_{1}|a_1,a_2,\alpha_1,\alpha_2)=\begin{cases} 1  {\ \rm if\ }  a_1=\text{sgn}(x)(\alpha_2)\\0 {\ \rm otherwise\,,}\end{cases}
\end{eqnarray}
one obtains a model that violates SI where, when conditioning on $\lambda=\lambda_1$, one can effectively set the value of $a_1$ at a distance, by setting the value of $\alpha_2$. In the language of DAGs, the paths $\alpha_1\!\to\!\lambda\!\leftarrow\!a_2$ and $\alpha_2\!\to\!\lambda\!\leftarrow\!a_1$ in Fig.\ 3  allow distant settings to affect local outcomes when we condition on the collider $\lambda$.

Imposing SI,  which involves imposing  the marginal distributions $P(a_1=\pm1 \mid \alpha_1)=\tfrac12$ and $P(a_2=\pm1\mid \alpha_2)=\tfrac12$, eliminates this possibility of superluminal signalling. Notably, under any $|\psi_{1;2;3;4}\rangle$ and for any fixed $\alpha_1$,
\begin{eqnarray}
P(a_1|\alpha_1,\alpha_2,\lambda_{1})=\sum_{a'_2}\,P_{|\psi_{1;2;3;4}\rangle}(a_1,a'_2|\alpha_1,\alpha_2)\nonumber\\
=1/2 = P(a_1|\alpha_1,\alpha'_2,\lambda_1)\,,\nonumber
\end{eqnarray}
showing that changing $\alpha_2$ leaves $a_1$'s distribution unaffected, and so no superluminal signals can be sent. An analogous argument holds for $a_2$. Hence, while the model violates LC, it is carefully fine-tuned via SI to respect the no-signaling constraint.

Violations of LC are widely believed to be in tension with the space-time structure of special relativity because they seem to clash with an intuitive idea of what it means for causal influence to travel within the light cone. The way in which the model proposed here violates LC seems benign in this respect: On the one hand, the backward-in-time law-like probabilistic dependences indicated by the arrows in Fig.\ 3 may be taken to be within the light cone. On the other hand, they cannot be exploited for actively manipulating the past due to the implementation of SI. A more detailed comparison between violations of LC in the present model and others, such as Bohmian mechanics, is beyond the scope of this letter.

\section{Accounting for GHZ correlations}

The approach from the previous section generalizes to systems with more than two components. In particular, it can recover the well-known GHZ correlations obtained from the state
\begin{equation}
|\psi\rangle_{\mathrm{GHZ}}
\;=\;
\frac{1}{\sqrt{2}}
\bigl(
  |1,1,1\rangle
  \;+\;
  |-1,-1,-1\rangle
\bigr),
\end{equation}
as originally introduced in \cite{ghz,ghsz,mermin}. To demonstrate this, we add a third measurement setting~\(\alpha_3\) and its associated outcome~\(a_3\) to the diagram from Fig.\,3 (not shown explicitly here). For simplicity, let each setting \(\alpha_i\) take values in \(\{0,1\}\), where \(\alpha_i=0\) denotes a measurement along the \(x\)-axis, and \(\alpha_i=1\) denotes a measurement along the \(y\)-axis. In these conventions, the GHZ correlations can be written as
\begin{eqnarray}
&&P_{|\psi\rangle_{GHZ}}(a_1,a_2,a_2|\alpha_1,\alpha_2,\alpha_3)\label{GHZcorrel}\\
&&\nonumber\hspace{1cm}=\begin{cases}\frac{1}{4} {\ \rm if\ } a_1a_2a_3=(-1)^{\alpha_1+\alpha_2+\alpha_3}\\0 {\ \rm otherwise\,.}\end{cases}
\end{eqnarray}
To reproduce these probabilities using a model of the form in Fig.\,3 (but with an extra wing for \(\alpha_3,a_3\)), we define
 \begin{eqnarray}
&\hspace{-3cm}P(\lambda_0|a_1,a_2,a_3,\alpha_1,\alpha_2,\alpha_3)\nonumber\\&=\mathcal N P_{|\psi\rangle_{GHZ}}(a_1,a_2,a_3|\alpha_1,\alpha_2,\alpha_3)\,,\label{recoverGHZ}
\end{eqnarray}
where \(\lambda_0\) denotes the hidden variable label for the GHZ state, and \(\mathcal{N} \equiv \tfrac{P(\lambda_0)}{P(a_1\mid \alpha_1)\,P(a_2\mid \alpha_2)\,P(a_3\mid \alpha_3)}\).

The GHZ state famously reveals a sharper contradiction with local realism than Bell's theorem alone, because there is no single assignment of \(\{x_1,y_1,x_2,y_2,x_3,y_3\}\) that satisfies all GHZ-type conditions. Specifically, one would need a configuration in which
\[
  x_1\,x_2\,x_3 \;=\; +1,
  \quad
  x_1\,y_2\,y_3 \;=\;
  y_1\,x_2\,y_3 \;=\;
  y_1\,y_2\,x_3 \;=\; -1,
\]
yet a simple multiplication of these equalities shows they cannot all hold simultaneously. This is often interpreted to mean that outcomes \(a_i\) in a GHZ experiment cannot be \emph{revealed} or ``pre-existing'' values of \(x_i\) or \(y_i\), since no classical hidden-variable assignment can fulfill all GHZ correlations at once.

In our model, by contrast, each outcome \(a_i\) \emph{is} identified with the relevant variable $x_i$ (if $\alpha_i=0$) or $y_i$ (if $\alpha_i=1$). The GHZ correlations \eqref{GHZcorrel} then hold only for the \emph{actual combination} of measured values in each run, and not for other possible combinations of $x_i$ or $y_i$. In effect, we ``allow'' the triple $(a_1,a_2,a_3)$ to determine the hidden variable's value:
\[
  \lambda = \lambda_0
  \quad
  \Leftrightarrow
  \quad
  (a_1,a_2,a_3)
  \text{ is GHZ-allowed.}
\]
Outcome triples that are not GHZ-allowed lead to $\lambda\neq \lambda_0$. Consequently, whenever the GHZ state $\lambda_0$ is prepared, any triple $(a_1,a_2,a_3)$ that violates the condition in \eqref{GHZcorrel} is simply \emph{not observed}.

Hence, the ``mechanism'' enabling these revealed values does not rely on multiple classical assignments being simultaneously valid for all measurement axes. Instead, the backward arrows into $\lambda$ let the model select \emph{one} appropriate triple of outcomes, consistent with the GHZ correlations, and thereby reproduce the quantum predictions in each run. In short, the model's backward dependency on $(a_1,a_2,a_3,\alpha_1,\alpha_2,\alpha_3)$ ensures that only outcome triples that match the GHZ pattern are realized under $\lambda_0$.

\section{Summary and outlook}

This paper has demonstrated how the predictions of maximally entangled states can be reproduced by models whose probability distributions factorize as indicated in Fig.~3. By imposing \emph{Statistical Independence} on these ``backward-arrow'' models, one can avoid superluminal signaling while still violating \emph{Local Causality} and thus recovering a Bell-CHSH inequality violation.

The models proposed here two notable limitations. First, they do not address why quantum correlations respect the Tsirelson bound. Although no quantum prediction can exceed $2\sqrt{2}$ on the left-hand side of Eq.~(\ref{inequality}), one can devise other non-signalling (but non-quantum) models, most famously the ``Popescu-Rohrlich box'' \cite{popescurohrlich}, which pushes this value to 4. The models introduced here are flexible enough to reproduce such extreme correlations, hence offering no insight into the deeper reason for the Tsirelson bound.

Second, the backward-arrow approach treats $\lambda$ purely as an \emph{effect} of later measurement variables and outcomes. In practice, $\lambda$ labels the prepared (Bell) state, which plausibly depends at least partially on earlier experimental actions. Consequently, it seems physically implausible to have only backward arrows into $\lambda$. More complete models should separate the collider variable $\lambda$ from a preparation variable $\mathcal P$ and assign $\lambda$ to a physical quantity capable of a suitable dynamical role.   It is to be expected that this would be independently necessary if one were to extend the approach proposed here to non-maximally entangled states, where the probability of the individual local outcome $a_i$ conditional on the associated setting $\alpha_i$ is no longer universally $\tfrac12$ but depends on the preparation procedure $\mathcal P$. This indicates that forward-in-time arrows from the preparation procedure $\mathcal P$ to the outcomes $a_i$ would be needed, making it necessary to distinguish between $\mathcal P$ and the collider variable $\lambda$.

An interesting candidate  generalization of the present model towards a fully-fledged ``single-world realist'' account of quantum theory is the Q-function-based model proposed by Drummond and coauthors \cite{drummond,drummondreid}, which likewise invokes backward-directed conditional probabilities but uses the positive semi-definite Husimi Q-function as a phase-space probability. As argued in \cite{friederich}, that model falls outside the popular \emph{ontological models framework} \cite{harriganspekkens} and hence is not ruled out by standard no-go theorems proved within that framework.

To conclude, this paper isolates, for the first time, a precise ingredient---namely, the systematic use of Statistical Independence as a fine-tuning condition---that renders backward-directed conditional probabilities both non-trivial and no-signaling. This might be seen as an important step toward a compelling ``single-world realist'' account of the quantum world while simultaneouslky being in harmony with space-time symmetries and recovering the EPR correlations.


\section*{Acknowledgments}
I am grateful to Huw Price, Ken Wharton, and the members of the Groningen Philosophy of Physics Group for helpful comments and discussions on an earlier version. This research was funded by the Netherlands Organization for Scientific Research (NWO), project VI.Vidi.211.088. GPT-4o and o1 were used for polishing the exposition.

\end{document}